# WGAN based Inverse Design of Active Dual Band FSS with Switchable Transmission

Rui Xi, Xinke Kuang, Huanran Qiu, Shiyun Ma, Xiaokui Kang, Yuanyuan Wang, Ying Li, Long Li.

***Abstract*—This letter presents a novel design method for switchable dual band transmissive frequency selective surface (FSS). The proposed FSS possesses characteristics of maintaining passband characteristics at high frequencies, while switching from transmission to reflection at low frequencies with pin diodes states altering. Specifically, we propose a crystal growth-based topology generation strategy, and utilize a simplified U-Net Wasserstein GAN (WGAN) neural network model to establish an inverse mapping model from electromagnetic response to structure topology parameters. The trained WGAN achieves training and validation accuracies of 95.59% and 90.84%, while the simplified U-Net attains training and validation accuracies of 98.5% and 94.1%. Using the trained WGAN. The generated structural topologies were validated through full-wave simulations and experimental measurements. The proposed method enhances the design flexibility and overcomes the time-consuming drawbacks of conventional FSS design.**

***Index Terms*— Frequency selective surface (FSS), active FSS, inverse design, Wasserstein generative adversarial networks (WGANs).**

## I. INTRODUCTION

In recent years, frequency selective surface (FSS) have emerged as a pivotal technology for wavefront manipulation across microwave to millimeter-wave frequencies by integrating PIN diodes, varactor diodes, or other active elements to enable reconfiguration of electromagnetic responses. Such structures enable multifunctional integration, including beam scanning, polarization conversion, and frequency selection [1],[2], with significant potential in 6G communications, radar detection, satellite systems, and intelligent reflecting surfaces [3]. Concurrently, numerous FSS-related studies have demonstrated their versatility, such as in fast beam-switching communication systems [5], dynamic radar interference for electronic warfare [6], and multi-beam programmable designs [7].

Conventional FSS designs predominantly rely on full-wave electromagnetic simulations coupled with global optimization algorithms, such as genetic algorithms (GAs) [8],[9] and particle swarm optimization (PSO) [10],[11], to refine unit geometries or phase distributions for desired responses. However, these methods often demand thousands of manual iterations, exhibit susceptibility to local optima, sluggish convergence, initial-value sensitivity, and surging computational burdens in high-dimensional cases like multi-state or multi-band configurations [12],[13].

Recently, generative adversarial networks (GANs) and their variants have been integrated into FSS inverse design [14],[15], adeptly mapping electromagnetic responses to structural parameters through adversarial training to boost efficiency [16]. Notably, Wasserstein GAN (WGAN) enhances stability via Wasserstein distance and gradient penalties, averting mode collapse and improving convergence in multifaceted tasks [17], enabling seconds-scale structure generation versus hours-to-days for GA/PSO. This approach has attracted growing interest, as seen in improved GANs for dual-functional designs and unsupervised methods for transmissive FSS synthesis [18].

In this letter, we propose a novel transmissive switchable FSS structure, leveraging WGAN for unit geometry optimization. This design achieves dual-band decoupled modulation, with adaptive transmission/rejection switching at low frequencies, while maintaining transmission at high frequencies to enhance multi-band system compatibility and radar-communication integration. Furthermore, the WGAN-driven non-parametric optimization yields efficient, non-intuitive unit geometries that surpass traditional parametric constraints, synergistically optimizing electromagnetic performance and structural simplicity. This approach offers a new paradigm for multi-band FSSs, dynamic beam steering, and intelligent electromagnetic environment engineering.

## II. DESIGN OF DATASETS AND NETWORK

### *A. Datasets design*

As shown in Fig. 1., the FSS unit cell is designed with dimensions of 3 × 3 $mm^2$ and comprises three metal layers printed on two F4B substrates, each substrate is with the dielectric constant of 3.5 and thickness of 1*mm*. The employed PIN diode model is MADP-000907-14020P. To ensure identical transverse electric (TE) and transverse magnetic (TM) dual-polarization responses, the entire unit is partitioned into four equivalent quadrants, each incorporating two regions designated for random generation. To promote sufficient diversity in the generated dataset, distinct random regions utilize differentiated pattern generation logics.

Corresponding author: Rui Xi, Ying Li, Long Li.

Rui Xi, Xinke Kuang, Huanran Qiu,Shiyun Ma, Xiaokui Kang and Yuanyuan Wang are with the Hangzhou Institute of Technology, Xidian University, Hangzhou 311231, China.

Ying Li is with the State Key Laboratory of Extreme Photonics and Instrumentation ZJU-Hangzhou Global Scientific and Technological Innovation Center, Zhejiang University, Hangzhou 310027, China

Long Li is with the Key Laboratory of High-Speed Circuit Design and EMC of Ministry of Education School of Electronic Engineering, Xidian University, Xi'an 710071, China

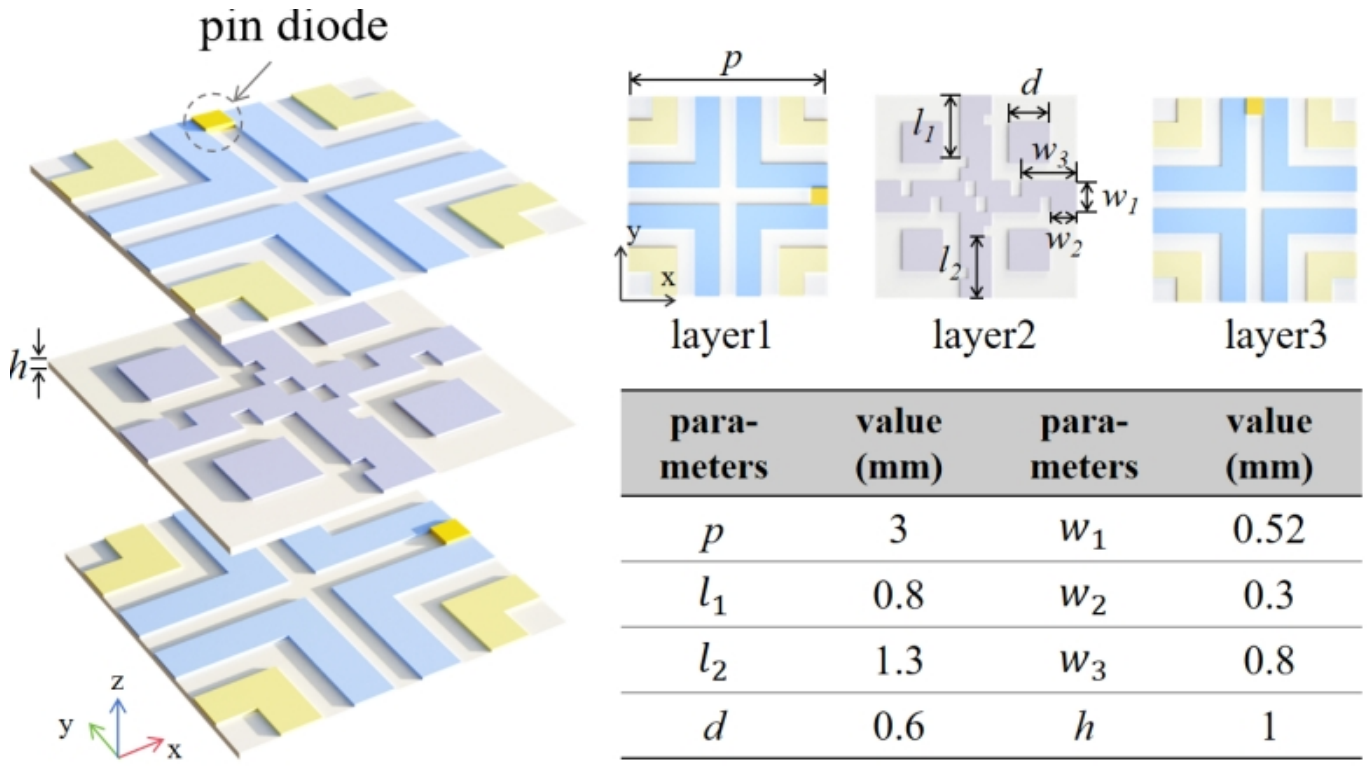

| para-meters | value (mm) | para-meters | value (mm) |
|---|---|---|---|
| $p$ | 3 | $w_1$ | 0.52 |
| $l_1$ | 0.8 | $w_2$ | 0.3 |
| $l_2$ | 1.3 | $w_3$ | 0.8 |
| $d$ | 0.6 | $h$ | 1 |

**Fig. 1.** The structure configuration of the proposed FSS

For the corner regions, a strategy involving three randomly selected points is adopted, where the maximum inter-point distance defines the radius of circles centered at each point; the resultant structure is formed by the intersection of these circular areas with the original region, thereby introducing partial variations while preserving the overall corner morphology.

At the same time, the central regions leverage a crystal growth mechanism: a point is randomly chosen along the edge as the initial crystal tip, and the surrounding area is iteratively evaluated for continued growth feasibility to maintain single connectivity. For permissible growth zones, new crystal tips are probabilistically attempted in a clockwise sequence, with the probability of generating a subsequent tip at the current location decrementing upon each successful extension. The fully developed crystal is subsequently subtracted from the central primitive region, yielding structures with substantial diversity yet comparable contours. The randomly generated central and corner regions are then concatenated to produce a single randomized structure per iteration.The complete random structure generation procedure is shown in Fig. 2.

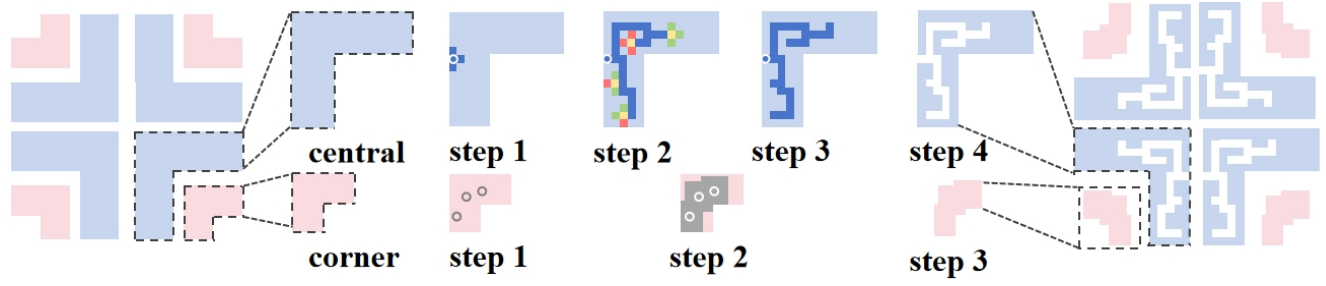

**Fig. 2.** Random structure generation procedure

*B. Predictor Design*

To supplant the computationally intensive full-wave electromagnetic simulations in real-time (where simulations of the structures herein typically exceed half an hour in HFSS), this work embeds a lightweight U-Net-style convolutional neural network as a predictor within the WGAN framework. This network takes as input a single-channel 2D geometric image of the unit cell structure and directly outputs the complex S21 response over the specified frequency band, thereby providing the generator with millisecond-level high-fidelity electromagnetic performance evaluations.

The predictor structure adopted in this letter is illustrated in Fig. 3. The input structure is first normalized to 16×16 via adaptive average pooling. It then advances to the encoder, undergoing three convolutional stages with output channel counts of 8, 16, and 32, respectively, ultimately reducing the feature map to dimensions 2×2×32. The decoder employs transposed convolutional structures with skip connections, progressively upsampling the feature maps and concatenating them with corresponding encoder layer features, finally restoring to a 16×16×12 feature map. Following flattening, two fully connected layers map to the output dimension, where 400=200×2 corresponds to the S21 responses for TE and TM modes across the 5-29 GHz band at 200 uniformly sampled frequency points under PIN diode ON/OFF states.

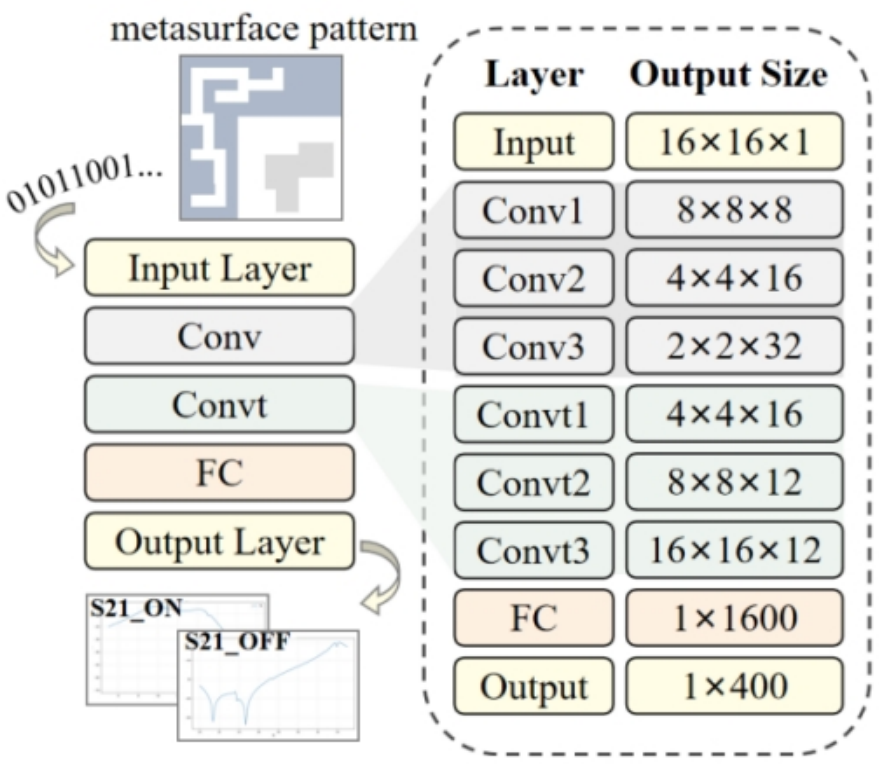

**Fig. 3.** Predictor structure

*C. Generative Adversarial Networks Design*

To achieve an end-to-end inverse mapping from target S21 responses to FSS unit cell structures, this letter designs a conditional generator based on a symmetric U-Net architecture. This generator accepts the desired S21 vector as conditional input and outputs a binary structural image of dimensions 14×14. Concurrently, a lightweight convolutional-multilayer perceptron (MLP) hybrid architecture is employed for the discriminator to enable mutual adversarial training between the discriminator and the generator.

The generator architecture is illustrated in Fig. 4(a). The input S21 vector is first mapped by two fully connected layers to a 14×14×64 feature map. It then passes through the U-Net encoder, which consists of two encoding stages — each compFSSing two convolutional layers and a dropout layer to prevent overfitting, followed by two intermediate convolutions, producing a 3 × 3 × 512 latent feature map. The decoder employs transposed convolutions with skip connections to upsample and fuse encoder features, ultimately yielding a 14×14 binary structure. This symmetric U-Net design preserves multi-scale spatial details, substantially enhancing the physical realizability and electromagnetic performance consistency of the generated structures.

The detailed discriminator structure is depicted in Fig. 4(b). The input structure is first processed through three convolutional layers to extract spatial features, resulting in a feature map of size 64×7×4, which is subsequently flattened into a one-dimensional feature vector of length 1792. This vector is concatenated with the S21 vector and fed into three fully connected layers, ultimately generating the discriminator's score for the input structure.

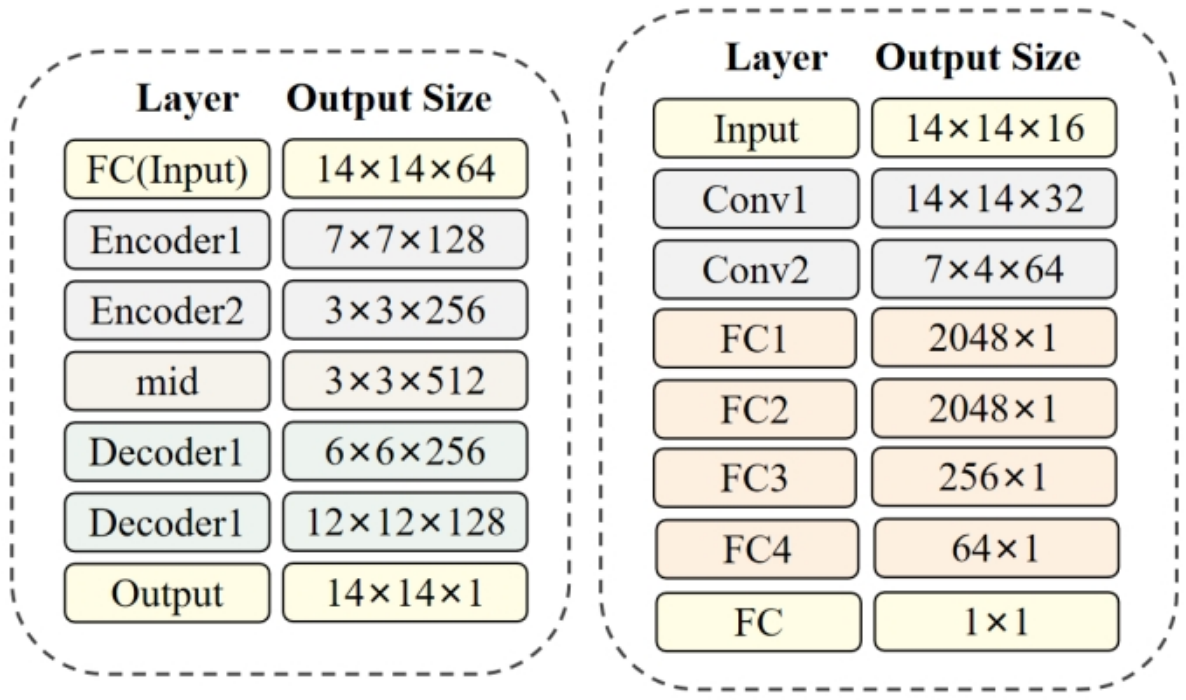

**Fig. 4.** Generative Adversarial Networks.

*D Loss Design*

The predictor is trained in a supervised manner, leveraging input structures from the dataset to learn inherent structural features and predict their S21 parameters, ultimately achieving surrogate effects for electromagnetic simulations. To enhance fitting accuracy in mid-frequency bands and effectively mitigate burrs in the predicted curves, this letter introduces a weighted hybrid loss function, as in

$$L_{Pred}=(1-\alpha)L_{weighted\text{-}MSE}+\alpha L_{FMAE},\quad \alpha=0.4 \quad (1)$$

$$L_{weighted\text{-}MSE}=\frac{1}{M}\sum_{i=1}^{M} w_i\left(\hat{S}_{21,i}-S_{21,i}\right)^2 \quad (2)$$

$$L_{FMAE}=\frac{1}{2}\left(\begin{array}{l}\frac{1}{K}\sum_{k=1}^{K}\left|\mathrm{Re}\{\mathrm{F}(\hat{\mathbf{s}}_k)-\mathrm{F}(\mathbf{s}_k)\}\right| \\ +\frac{1}{K}\sum_{k=1}^{K}\left|\mathrm{Im}\{\mathrm{F}(\hat{\mathbf{s}}_k)-\mathrm{F}(\mathbf{s}_k)\}\right|\end{array}\right) \quad (3)$$

In (2), the weighted mean squared error (MSE) assigns elevated weights to the centers of each frequency band, with the weight vector linearly ascending from 1.0 at the edges to 1.5 at the center, symmetrically mirrored, concatenated across bands, thereby facilitating network emphasis on core frequency segments. In (3), the frequency-domain mean absolute error component applies a real-valued fast Fourier transform to the output, computing the mean absolute error across the real and imaginary spectra to substantially attenuate spectral oscillations and burrs.

To ensure that the structures output by the generator concurrently fulfill the target S21 responses while exhibiting physical realizability, this work devises a hybrid loss function tailored specifically for the generator. This loss function supports both the WGAN warm-up phase and the formal adversarial training phase, amalgamating adversarial signals with multi-dimensional structural supervisory losses to ascertain that the generated structures are highly authentic and reliable in terms of both electromagnetic performance and geometric morphology.

The total generator loss is defined as

$$L_{Gen}=L_{struct}-\lambda_{adv}\cdot(1-\delta_{warmup})\cdot score \quad (4)$$

where $L_{struct} = L_{BCE} + L_{Dice} + L_{bitwise}$ denotes the structural supervisory loss, $\delta_{warmup}$ represents the warm-up indicator (set to 1 during the initial 20 epochs, wherein solely adversarial training is conducted to stabilize the discriminator and prevent misguidance of the generator by an inadequately trained discriminator), score signifies the discriminator's output score for the generated structure (in Wasserstein form, with larger values indicating greater "realism"), and $\lambda_{adv}$ is the adversarial weight.

The structural supervisory loss consists of the following three components:

$$L_{struct}=L_{BCE}+L_{Dice}+L_{bitwise} \quad (5)$$

BCE structural loss (mixture of soft and hard labels):

$$L_{BCE}=(1-\lambda)\,L_{BCE}^{soft}+\lambda\,L_{BCE}^{hard},\ \lambda=0.5 \quad (6)$$

Where $L_{BCE}^{soft}$ denotes the binary cross-entropy with logits between the output and the target boundary distance map, and $L_{BCE}^{hard}$ denotes the binary cross-entropy with logits between the flattened output and the flattened target binary map, respectively.

Dice loss (to enhance foreground region overlap):

$$L_{Dice}=1-\frac{2\sum(\hat{y}\odot y)+\varepsilon}{\sum(\hat{y}+y)+\varepsilon},\ \varepsilon=10^{-6} \quad (7)$$

where $\hat{y}$ represents the sigmoid-activated output of the generator, $y$ is the target binary map, $\odot$ denotes element-wise multiplication.

Bitwise pixel-level matching loss:

$$L_{bitwise}=1-\frac{1}{BHW}\sum_{b=1}^{B}\sum_{h=1}^{H}\sum_{w=1}^{W} I\left(\sigma(\hat{y}_{b,h,w})=y_{b,h,w}\right) \quad (8)$$

where $\sigma(\cdot)$ is the sigmoid function, $\mathbb{I}(\cdot)$ is the indicator function , and $B, H, W$ are the batch size, height, and width, respectively.

## III. WORKFLOW AND TRAINING RESULTS

*A. Workflow*

Fig. 5 illustrates the WGAN-based structure generation workflow. Random structures are first generated in Python and automatically simulated in HFSS, producing a dataset of 2000 samples. The dataset is subsequently used to train the predictor under supervised learning with an early-stopping mechanism: training terminates if the loss exhibits no sustained decline over 20 consecutive epochs, thereby preventing overfitting. Upon completion of predictor training, the dataset is fed into the WGAN for adversarial training. To prevent early misguidance and convergence to local optima, the generator and discriminator are decoupled during the initial phase—the discriminator receives generator outputs, while the generator is temporarily withheld from discriminator scores. An independent generator loss compFSSing bitwise and BCE components is simultaneously monitored as an additional early-stopping criterion. After the prescribed training epochs or early stopping, the target response curve is input to the generator; the resulting output serves as the WGAN-generated structure and is promptly validated for electromagnetic performance by the predictor.

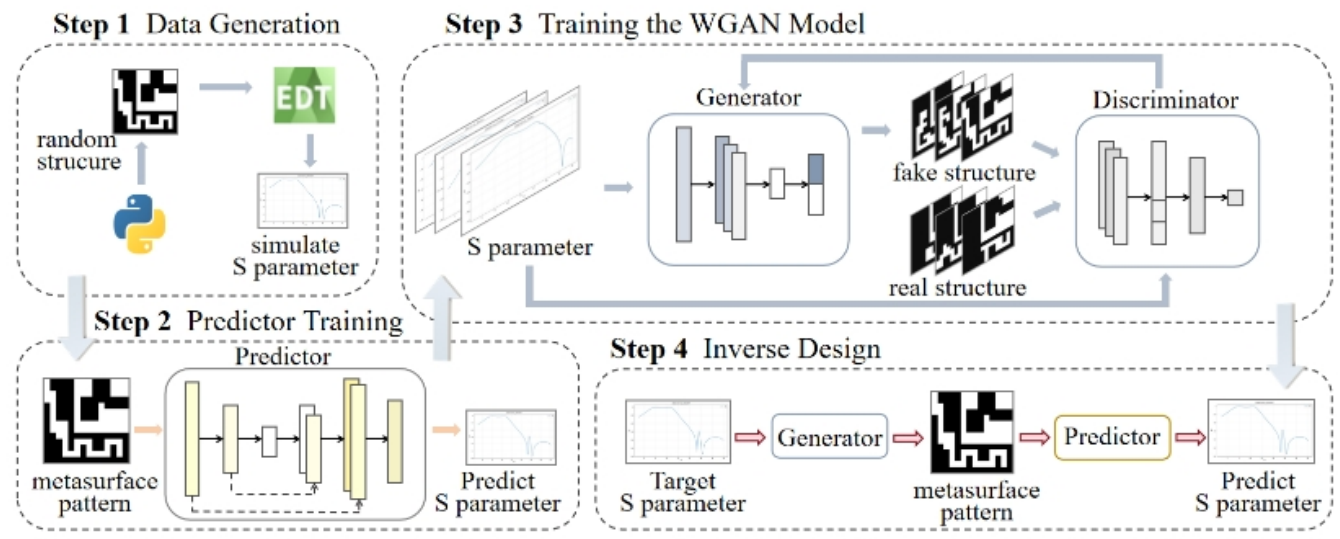

**Fig. 5.** Workflow of the WGAN-based architecture for struct generate

*B. Training results*

Following training on a dataset comprising 2000 samples, the network loss convergence curves are depicted in Fig. 6(a) and (b), where (a) illustrates the predictor's training loss curve and (b) shows the generator's independent loss curve. Upon training completion, the predictor achieves a validation set loss of 0.062, while the generator attains a validation set loss of 0.069. Figs. 6(c) and (d) demonstrate the post-training performance of the predictor on the validation set.

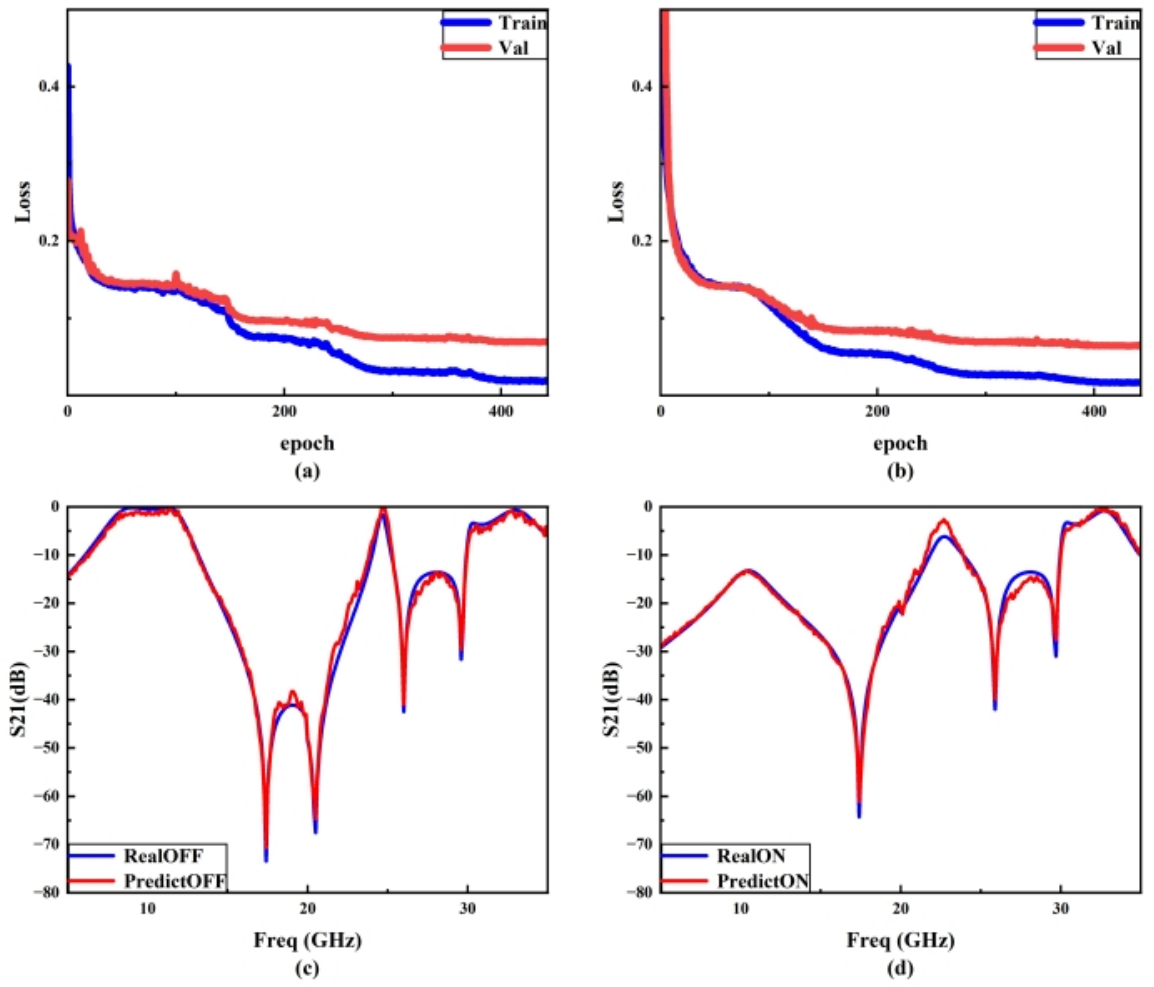

**Fig. 6.** (a) Predictor's training loss. (b) Generator's training loss. (c),(d) Predictor validation performance

The target S21 response curve is defined such that it exhibits switchable charactefsstics in the low-frequency band (6-10 GHz) and maintains a constant passband in the high-frequency band (23-29 GHz). Within the passband regions, the transmission coefficients are randomly selected from 0 to -2 dB, whereas those in the remaining frequency segments are randomly chosen from -50 to -60 dB. This target curve is input to the generator network to obtain generated structures, with their electromagnetic responses rapidly evaluated via the predictor. The process is repeated 50 times to yield 50 candidate structures. Subsequently, screening is performed based on the predictor's output electromagnetic responses, ultimately selecting the structure that meets the design requirements, as illustrated in Fig. 7. Full-wave electromagnetic simulations of the structure demonstrate switchable charactefsstics across the 7.67-11.7 GHz band and a constant passband in the 24.3-30.5 GHz band, exhibiting close agreement with the target response curve.

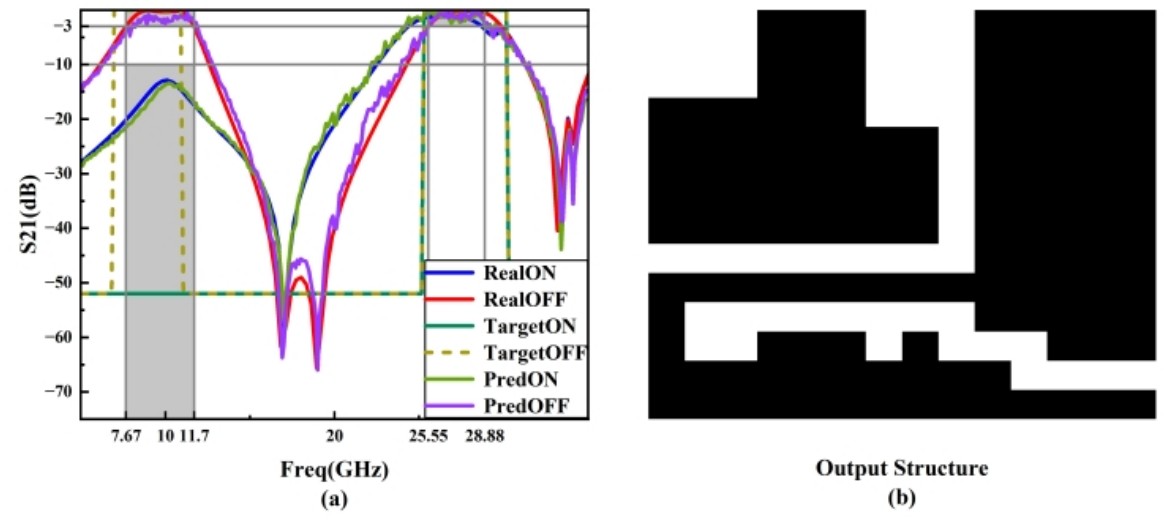

**Fig. 7.** (a) Comparison of Target, Predicted, and Measured S21 Responses. (b) The Generator output structure.

## IV. EXPERIMENTAL VERIFICATION

To validate the electromagnetic performance of the generated structure, an 18×18 array prototype was fabricated and measured. The prototype employs 1-mm-thick F4B substrates with relative permittivity 3.5, where the presence or absence of metal patches emulates the ON/OFF states of the PIN diodes. As shown in Fig. 8 (a) and (b), measurements were performed using the waveguide method with ports operating in the 6-10 GHz, 10-15 GHz, and 20-33 GHz bands; the obtained data were spliced to produce the complete response, as depicted in Fig. 8 (c) and (d). The measured results indicate a low-frequency operating band of 8-11.7 GHz and a high-frequency operating band of 24.3-30.5 GHz, demonstrating good agreement with the full-wave simulation results.

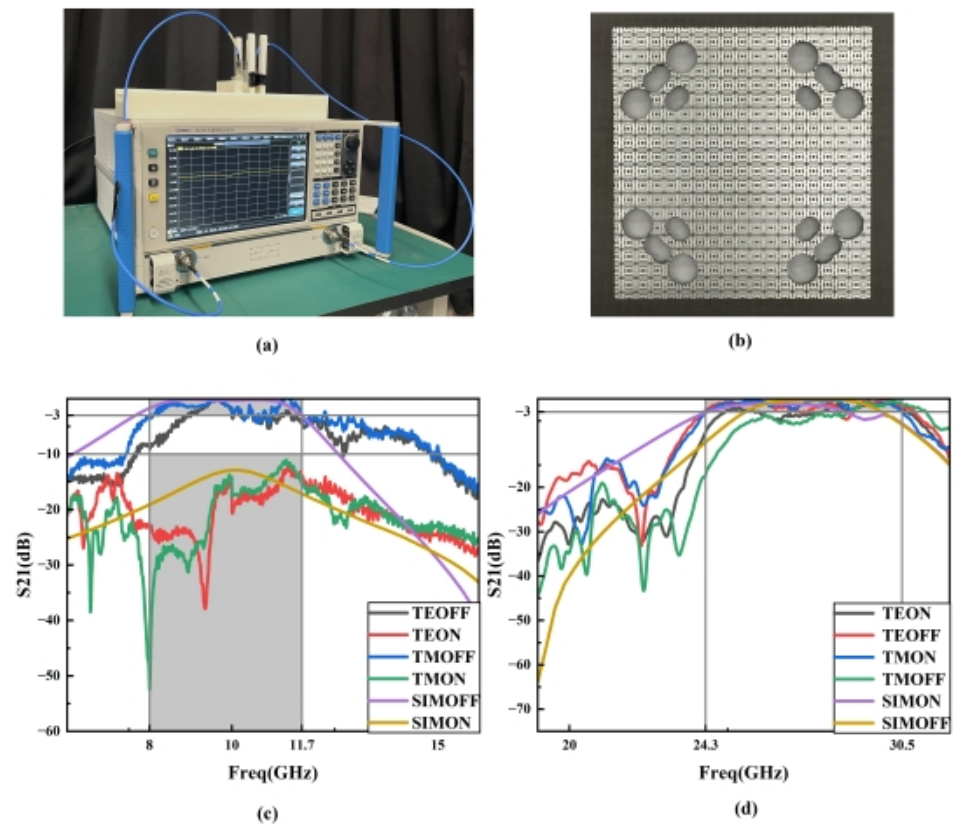

**Fig. 8.** (a),(b) Measurement setup. (c),(d) Measured transmission coefficient.

## V. CONCLUSION

This letter presents a WGAN-based method for the inverse design of FSSs. A lightweight predictor is developed to provide rapid real-time prediction of electromagnetic responses, replacing computationally intensive full-wave simulations, while an improved-loss-function WGAN is employed for efficient inverse structure optimization. The generated structures are comprehensively validated through predictor evaluation, full-wave simulations, and experimental measurements on a prototype, demonstrating the effectiveness and superiority of the proposed approach. Future work will focus primarily on further refinement of the network architecture.